\documentstyle[preprint,aps]{revtex}
\begin{document}
\draft

\title
{Classical Dirac Observables: the Emergence of Rest-Frame Particle and Field
Theories}

\author{Luca Lusanna}

\address
{Sezione INFN di Firenze\\
L.go E.Fermi 2 (Arcetri)\\
50125 Firenze, Italy\\
E-mail LUSANNA@FI.INFN.IT}

\maketitle
 
\vskip 1truecm
\noindent Talk at the ``Pacific Conference on Gravitation and Cosmology",
1-6 February 1996, Seoul, Korea

The second Noether theorem\cite{noe} and Dirac-Bergmann constraint theory
\cite{di,be} are the basis respectively of the Lagrangian and Hamiltonian
formulations of all relativistic physical systems\cite{lus}. The need of
redundant variables, to be reduced due to the presence of either first
and/or second class constraints, is connected with requirements like
manifest covariance and minimal coupling (the gauge principle). Also Newton
mechanics\cite{nm} and Newton gravity with Galilean general covariance
\cite{ng} can be reformulated in this language at the nonrelativistic level.
Therefore, at the Hamiltonian level the fundamental geometric structure
behind our description of physics is presymplectic geometry\cite{lich,go},
the theory of a closed degenerate two-form (no definition of Poisson Brackets). 
Curiously, it has been much less studied than its dual structure, Poisson 
geometry, namely the theory of a closed (i.e. with vanishing 
Schouten-Nijenhuis bracket with itself)  degenerate bivector (existence of
degenerate Poisson brackets)\cite{lic}; in absence of degeneracy, Poisson
manifolds coincide with symplectic manifolds.

It is important to understand the properties of presymplectic manifolds
embedded into ambient phase spaces, so to utilize their natural Poisson
brackets as in the Dirac-Bergmann theory. In particular, for physical
applications, Darboux charts of presymplectic manifolds are needed, for
instance in the definition of the Faddeev-Popov measure in the path integral.
The tool to find such charts are the Shanmugadhasan canonical transformations
\cite{sha} and the multitemporal version of the equations defining gauge
transformations\cite{lum}.

By restricting ourselves to systems with only first class constraints (most
of relevant physical systems are of this type), one looks for new canonical
bases in which all first class constraints are replaced by a subset of the new
momenta (Abelianization of the constraints); then the conjugate canonical
variables are  Abelianized gauge variables and the remaining canonical pairs
are special Dirac observables in strong involution with both Abelian
constraints and gauge  variables. These Dirac observables, together with the
Abelian gauge variables, form a local Darboux system of coordinates for the
presymplectic manifold $\bar \gamma$ defined by the original first class
constraints (this manifold is coisotropically embedded\cite{go} in the
original phase space, if suitable mathematical conditions are satisfied).
In the multi-temporal method each first class constraint
is raised to the status of a Hamiltonian with a time-like parameter
describing the associated evolution (the genuine time describes the evolution
generated by the canonical Hamiltonian, after extraction from it of the
secondary and higher order first class constraints): in the Abelianized form
of the constraints these "times" coincide with the Abelian gauge variables on
the solutions of the Hamilton equations. These coupled Hamilton equations are
the multi-temporal equations: their solution describes the dependence of the
original canonical variables on the time and on the parameters of the
infinitesimal gauge transformations, generated by the first class constraints.
Given an initial point on the constraint manifold, the general solution
describes the gauge orbit, spanned by the gauge variables, through that point;
instead the time evolution generated by the canonical Hamiltonian (a first
class quantity) maps one gauge orbit into another one. For each system the
main problems are whether the constraint set is a manifold (a stratified
manifold, a manifold with singularities...), whether the gauge orbits can be
built in the large starting from infinitesimal gauge transformations and
whether the foliation of the constraint manifold (of each stratum of it) is
either regular or singular. Once these problems are understood, one can check
whether the reduced phase space (Hamiltonian orbit space) is well defined.

Since for all isolated systems defined on Minkowski spacetime there is the
Poincar\'e kinematical symmetry group globally canonically implemented
\cite{pau} [for field theories the boundary conditions on the fields must be 
such that the ten Poincar\'e generators are finite], the presymplectic manifold 
$\bar \gamma$ is a stratified manifold with the main stratum (dense in $\bar 
\gamma$) containing all configurations belonging to timelike Poincar\'e
realizations with spin [$P^2 > 0$, $W^2=-P^2{\hat {\vec S}}^2\not= 0$; ${\hat
{\vec S}}$ is the rest-frame Thomas spin]. Then there will be strata with
i) $P^2 > 0$, $W^2=0$, and ii) $P^2=0$; the spacelike stratum $P^2 < 0$ must be
absent, otherwise there would be configurations of the system violating
Einstein causality. Each stratum may have further stratifications and/or 
singularity structures according to the nature of the physical system.

Therefore, the canonical bases best adapted to each physical system will be the
subset of Shanmugadhasan bases which, for each Poincar\'e stratum, is also
adapted to the geometry of the corresponding Poincar\'e orbits with their
little groups. These special bases could be named Poincar\'e-Shanmugadhasan
(PS) bases for the given Poincar\'e stratum of the presymplectic manifold;
till now only the main stratum $P^2 > 0$, $W^2\not= 0$, has been investigated.
Usually PS bases are defined only locally and one needs an atlas of these charts
to cover the given Poincar\'e stratum of $\bar \gamma$; for instance this
always happens with compact phase spaces.

When the main stratum of a noncompact physical system admits a set of global
PS bases (i.e. atlases with only one chart), we get a global symplectic
decoupling (a strong form of Hamiltonian reduction) of the gauge degrees of
freedom from the physical Dirac observables without introducing gauge-fixing
constraints; this means that the global PS bases give coordinatizations of the
reduced phase space (the space of Hamiltonian gauge orbits or symplectic
moduli space).

The program of symplectic decoupling was initiated by Dirac himself\cite{dir},
who found the Dirac observables of the system composed by the electromagnetic
field and by a fermionic (Grassmann-valued) field. After the rediscovering
of this method in the study\cite{longhi} of the relativistic two-body
DrozVincent-Todorov-Komar model\cite{dv,todo,komar}, it was applied to the
Nambu string\cite{colomo}. Then, the Dirac observables of Yang-Mills theory
with fermion fields were found\cite{lusa} in the case of a trivial principal
bundle over a fixed-$x^o$ $R^3$ slice of Minkowski spacetime with suitable
Hamiltonian-oriented boundary conditions. After a discussion of the
Hamiltonian formulation of Yang-Mills theory, of its group of gauge
transformations and of the Gribov ambiguity, the theory has been studied in
suitable  weigthed Sobolev spaces where the Gribov ambiguity is absent.
The physical Hamiltonian has been obtained: it is nonlocal but without any
kind of singularities, it has the correct Abelian limit if the structure 
constants are turned off, and it contains the explicit realization of the 
abstract Mitter-Viallet metric.
Subsequently, the Dirac observables of the Abelian and non-Abelian SU(2)
Higgs models with fermion fields were found\cite{lv1,lv2}, where the
symplectic decoupling is a refinement of the concept of unitary gauge.
There is an ambiguity in the solutions of the Gauss law constraints, which
reflects the existence of disjoint sectors of solutions of the Euler-Lagrange
equations of Higgs models. The physical Hamiltonian and Lagrangian of  the
Higgs phase have been found; the self-energy turns out to be local and
contains a local four-fermion interaction. It is now in preparation a
paper\cite{lv3} on the Dirac observables of the standard SU(3)xSU(2)xU(1) model
of elementary particles, using the previous results.

However, all these Hamiltonian reductions of gauge field theories suffer
of the problem of Lorentz covariance: one cannot make a complete
Hamiltonian reduction for systems defined in Minkowski spacetime without a
breaking of manifest Lorentz covariance. A universal solution of this probblem
has been found by reformulating\cite{lusa,lus1,karp}  every relativistic system
on a family of spacelike hypersurfaces foliating Minkowski spacetime\cite{di}.

As shown in these papers, in this way way one obtains the minimal breaking of 
Lorentz covariance: after the restriction from arbitrary spacelike
hypersusrfaces to spacelike hyperplanes, one selects all the configurations
belonging to the main Poincar\'e stratum and then restricts oneself to
the special family of hyperplanes orthogonal to the total momentum of the
given configuration (this family may be called the Wigner foliation of
Minkowski spacetime intrinsically defined by the given system).
In this way only three physical degrees of freedom,
describing the canonical center-of-mass 3-position of the overall isolated
system, break Lorentz covariance, while all the field variables are either
Lorentz scalars or Wigner spin-1 3-vectors transforming under Wigner
rotations. This method is based on canonical realizations of the Poincar\'e
group on spaces of functions on phase spaces\cite{pau} 
and one has the transposition at 
the canonical level of the techniques used to study the irreducible
representations of the Poincar\'e group and the relativistic wave equations.

Therefore one has to study the problem of the center of mass of
extended relativistic systems in irreducible representations of the Poincar\'e
group with $P^2 > 0$, $W^2=-P^2{\vec {\bar S}}^2\not= 0$ : it can be shown
that this problem leads to the identification of a finite world-tube of
non-covariance of the canonical center-of-mass, whose radius $\rho =\sqrt{-W^2}
/P^2=|\, {\vec {\bar S}}\, |\, /\sqrt{P^2}$ 
\cite{mol} identifies a classical intrinsic
unit of length, which can be used as a ultraviolet cutoff at the quantum
level in the spirit of Dirac and Yukawa [it exists also in asymptotically flat
general relativity due to the existence of the asymptotic Poincar\'e charges]. 
As mentioned in the papers\cite{lusa,lus1}, the distances
corresponding to the interior of the world-tube are connected with problems
coming from both quantum theory and general relativity: 1) pair production 
happens when trying to localize particles at these distances; 2) relativistic
bodies with a material radius less than $\rho$ cannot have the classical
energy density definite positive everywhere in every reference frame and the
peripheral rotation velocity may be higher than the velocity of light.
Therefore, the world-tube also is the flat remnant of the energy  conditions of
general relativity.

As shown in the paper\cite{lus1}, the formulation on spacelike hypersusrfaces
is also needed to solve the kinematics of N free scalar relativistic
particles described by N mass-shell first class constraints. While in the
standard formulation one has a description with N times, on the Wigner
hyperplane one gets a covariant 1-time (the Lorentz-invariant rest-frame
time) description: this new form of the dynamics (the ``rest-frame covariant
instant form" in Dirac's terminology\cite{diracc}) realizes in a covariant way
the separation of the canonical noncovariant center of mass from the relative
variables (with Wigner covariance) for every extended relativistic system
(particles, strings, classical field configurations). For particles it gives a
kinematical descrption free of the relative-times and relative-energies
variables [they are the sources of troubles in the theory of relativistic
bound states\cite{saz}; the basic problem is that Fock space, needed to 
introduce the particle concept in relativistic quantum field theory, is not a 
Cauchy problem for this theory: its asymptotic states are tensor
products of free particles and there is no restriction on their relative
times, so that some of them can be in the absolute future of the others] and
requires a choice of the sign of the energy for each particle [the intersection
of a timelike worldline with a spacelike hypersurface is determined by three
numbers]. Also an intrinsic covariant 1-time formulation of relativistic
statistical mechanics is possible in this framework\cite{lus1}. Finally the
formulation on spacelike hypersurfaces is the only known covariant way
of describing consistently the isolated system of N scalar charged relativistic
particles plus the electromagnetic field [i.e. of having a covariant closure
of the Poisson algebra of first class constraints]. When the Dirac observables
with respect to electromagnetic gauge transformations are worked out for
this system\cite{lus1}, one finds in the physical Hamiltonian the interparticle
Coulomb potential extracted from field theory and, if the electric charges
are described by Grassmann variables\cite{casal} [hypothesis of quantization
of the electric charge], one gets a regularized classical self-energy. It is
now under investigation along these lines the interaction of scalar particles
with Yang-Mills and linearized spin-2 fields in connection with the problems
of confinement and of finding a relativistic form of the Newton potential
[it would allow an exact evaluation of relativistic recoil effects in binary
systems] respectively; also one has to introduce the spin with Grassmann 
variables\cite{casal} to find the classical basis of the hydrogen atom and of 
positronium.

The rest-frame theory on Wigner hyperplanes has still an open problem,
namely in it there are four first class constraints one of which gives the
invariant mass of the system [it is the analogue of the Hamiltonian for
the relative motion in Newton mechanics]. The other three say that the total
three-momentum on the hyperplane (the intrisic rest frame) vanishes; but this
implies that one has still to eliminate the three conjugate gauge variables in 
the mass-shell constraint, so to remain only with it expressed in terms of
the final Dirac observables. With only particles this elimination is done
easily, but when fields are present the problem is still unsolved. It is now
under study how to define a canonical transformation realizing the
center-of-mass and relative-variables decomposition of a configuration of a
classical linear field theory. When this will be accomplished, there will be the
possibility to define a ``rest-frame quantum field theory" with suitable
Tomonaga-Schwinger asymptotic states (the true Cauchy problem for relativistic
field theory), to formulate a consistent bound-state theory and to attack in 
a new way the problem of how to introduce bound states among the asymptotic
states.

Finally, it is in an advanced stage the solution of 13 of the 14 first class
constraints of tetrad gravity in the asymptotically flat case. Here the aim
is to get the canonical transformation Abelianizing them in some suitable
system of coordinates and to find the form of the super-Hamiltonian constraint
in terms of the reduced Dirac's observables (giving a parametrization of the
superspace of three-geometries). If this can be done for some special family
of three-manifolds, then one can start with the attempt to put the four
interactions together, to find the final Dirac's observables of the standard
model coupled to tetrad gravity and to start with a quantization program 
with a physical cutoff.

\end{document}